\documentclass[twocolumn]{revtex4}
\usepackage{epsfig}
\usepackage{amssymb}
\usepackage{amsmath}
\usepackage{amsfonts}

\newcommand{\R}{\mathbb{R}}
\newcommand{\C}{\mathbb{C}}

\newcommand{\be}{\begin{equation}}
\newcommand{\ee}{\end{equation}}
\newcommand{\bea}{\begin{eqnarray}}
\newcommand{\eea}{\end{eqnarray}}
\newcommand{\nn}{\nonumber}
\newcommand{\kt}{\rangle}
\newcommand{\br}{\langle}

\newcommand{\ed}{\end{document}}

\newcommand{\pbr}{\prec\!}
\newcommand{\pkt}{\!\succ}

\newcommand{\bi}{\begin{itemize}}
\newcommand{\ei}{\end{itemize}}

\begin{document}
\title{Quantum Brachistochrone Problem and the Geometry of the State
Space in Pseudo-Hermitian Quantum Mechanics}

\author{Ali~Mostafazadeh}
\address{Department of Mathematics, Ko\c{c} University, Sariyer 34450,
Istanbul, Turkey\\
amostafazadeh@ku.edu.tr}

\begin{abstract}
A non-Hermitian operator with a real spectrum and a complete set
of eigenvectors may serve as the Hamiltonian operator for a
unitary quantum system provided that one makes an appropriate
choice for the defining inner product of the physical Hilbert
state. We study the consequences of such a choice for the
representation of states in terms of projection operators and the
geometry of the state space. This allows for a careful treatment
of the quantum Brachistochrone problem and shows that it is indeed
impossible to achieve faster unitary evolutions using ${\cal
PT}$-symmetric or other non-Hermitian Hamiltonians than those
given by Hermitian Hamiltonians.
\medskip

\hspace{6.1cm}{Pacs numbers: 03.65.-Xp, 03.67.Lx, 02.30.Yy,
02.40.-k}
\end{abstract}

\maketitle

Since the publication of the pioneering work of Bender and
Boettcher \cite{bender-prl-1998} on non-Hermitian ${\cal
PT}$-symmetric Hamiltonian operators, there have appeared numerous
research articles exploring the mathematical properties of such
operators and their possible physical applications. Recently, it
has been suggested that one can obtain arbitrarily fast quantum
evolutions using a class of such Hamiltonians
\cite{bender-prl-2007}. If true, this will have drastic
consequences in quantum computation, because for example it
removes the bound on the time-optimal unitary NOT operations
\cite{nielson-pra-2006} that is obtained within the framework of
conventional (Hermitian) quantum mechanics \cite{Vaidman,
Margolus,carlini,brody-hook}. As pointed out in \cite{martin},
this seems to contradict the equivalence of the quantum theory
based on such non-Hermitian Hamiltonians and the Hermitian quantum
mechanics \cite{jpa-2003,cjp-2004b}. In this article, we offer a
comprehensive treatment of this problem that is based on a
detailed study of the projective space ${\cal P H}_{\rm phys}$ of
physical states. In particular, we obtain the explicit form of the
natural metric tensor on ${\cal P H}_{\rm phys}$ and unravel the
subtleties of the quantum Brachistochrone problem for a general
unitary quantum system that is defined by a non-Hermitian
Hamiltonian.

In general if a linear (possibly non-Hermitian) operator has a
complete set of eigenvectors and a real spectrum, then it can
serve as the Hamiltonian operator for a unitary quantum system
provided that the physical Hilbert space of the system is defined
using an appropriate inner product
\cite{p2-p3,bender-prl-2002,jmp-2003}. This leads to a quantum
theory that turns out to be equivalent to the conventional quantum
mechanics \cite{jpa-2003,cjp-2004b}. In other words, this theory,
that we refer to as \emph{Pseudo-Hermitian Quantum Mechanics}
\cite{jpa-2004b}, is an alternative representation of the
conventional quantum mechanics. The key ingredient of this
representation is that the inner product of the physical Hilbert
space ${\cal H}_{\rm phys}$ is determined by the Hamiltonian
operator of the system. This has led to the discovery of an
intriguing structural similarity between quantum mechanics and
general theory of relativity \cite{pla-2004-jmp-2006a}. It has
also found applications in dealing with the Hilbert-space problem
in quantum cosmology \cite{ap-2004}, the old problem of
constructing a unitary first-quantized quantum theory of
Klein-Gordon fields \cite{ap-2006}, bound state scattering
\cite{matzkin}, and ghosts in certain quantum field theories
\cite{bender-review}.

In \cite{bender-prl-2007} the authors consider a class of
two-level non-Hermitian ${\cal PT}$-symmetric Hamiltonians, define
${\cal H}_{\rm phys}$ using the so-called ${\cal CPT}$-inner
product, and explore the evolution of state vectors, i.e.,
elements of ${\cal H}_{\rm phys}$. They conclude that for a fixed
initial and final state vectors, $\psi_I$ and $\psi_F$, one can
obtain a Hamiltonian operator that evolves $\psi_I$ into $\psi_F$
in an arbitrarily short time $\tau$. In this article we reconsider
this problem from the point of view that the true dynamics of
physical states occurs in the projective Hilbert space ${\cal P
H}_{\rm phys}$, i.e., the state space of rays in the physical
Hilbert space; a proper treatment of the quantum Brachistochrone
problem requires a closer look at ${\cal P H}_{\rm phys}$ and its
geometry.

In Hermitian quantum mechanics, the Hilbert space ${\cal H}$ is
defined by the usual $L^2$-inner product and the projective
Hilbert space ${\cal PH}$ is the projective space $\C P^{N-1}$
where $N$($\leq\infty$) is the dimension of ${\cal H}$. As pointed
out in \cite{brody-hook}, the lower bound on the duration $\tau$
of the evolution is proportional to the geodesic distance (between
the initial and final states) that is determined using the
Fubini-Study metric tensor on $\C P^{N-1}$,
\cite{anandan-aharonov}. This suggests constructing the analogous
metric tensor on ${\cal P H}_{\rm phys}$. A convenient method of
doing this is to represent the states (elements of ${\cal P
H}_{\rm phys}$) by appropriate projection operators and use the
relevant inner product on the space of linear (trace-class)
operators acting in ${\cal H}_{\rm phys}$ to induce the desired
metric tensor on ${\cal P H}_{\rm phys}$.

Let ${\cal H}$ be a Hilbert space defined by the conventional
$L^2$-inner product $\br\cdot|\cdot\kt$ and $H:{\cal H}\to{\cal
H}$ be a linear diagonalizable operator \footnote{This means that
$H$ has a complete set of eigenvectors. For a mathematically more
rigorous treatment see \cite{cjp-2006}.} with a real discrete
spectrum. Then there is a (positive-definite) inner product
$\pbr\cdot,\cdot\pkt$ that renders $H$ self-adjoint, i.e.,
$\pbr\psi,H\phi\pkt=\pbr H\psi,\phi\pkt$, \cite{p2-p3}. This inner
product is not unique but has the general form
$\pbr\cdot,\cdot\pkt=\br\cdot|\eta_+\cdot\kt$, \footnote{The
${\cal CPT}$-inner products \cite{bender-prl-2002} correspond to a
special class of the inner products $\pbr\cdot,\cdot\pkt$,
\cite{jmp-2003}.}, where $\eta_+:{\cal H}\to{\cal H}$ is a
positive-definite operator satisfying the pseudo-Hermiticity
condition: $H^\dagger=\eta_+H\eta_+^{-1}$, \cite{p1,jmp-2003}.
Here and throughout this paper we define the adjoint $A^\dagger$
of every linear operator $A$ using the $L^2$-inner product
$\br\cdot|\cdot\kt$ through the condition $\br\psi|A\phi\kt=\br
A^\dagger\psi|\phi\kt$. We say that $A$ is \emph{Hermitian} if
$A^\dagger=A$. We also introduce the $\eta_+$-pseudo-adjoint
$A^\sharp$ of $A$ that is defined by
$A^\sharp:=\eta_+^{-1}A^\dagger\eta_+$, \cite{p1}. This allows us
to express the $\eta_+$-pseudo-Hermiticity of $A$, i.e., the
condition $A^\dagger=\eta_+A\eta_+^{-1}$, as $A^\sharp=A$.

The physical Hilbert space ${\cal H}_{\rm phys}$ of the system
whose dynamics is determined by the Hamiltonian operator $H$ is
obtained by taking the linear span of the eigenvectors of $H$ in
${\cal H}$, endowing it with the inner product
$\pbr\cdot,\cdot\pkt$ for some (metric operator) $\eta_+$, and
completing the resultant inner product space. Clearly any linear
operator $A$ acting in ${\cal H}_{\rm phys}$ is self-adjoint if
and only if $A^\sharp=A$. These operators constitute the physical
observables of the system \cite{jpa-2004b}, the primary example
being the Hamiltonian $H$.

The Hilbert space ${\cal H}_{\rm phys}$ and the Hamiltonian $H$
define a pseudo-Hermitian quantum system that can be equivalently
described by the standard Hilbert space ${\cal H}$ and the
Hermitian Hamiltonian
$h:=\eta_+^{\frac{1}{2}}H\eta_+^{-\frac{1}{2}}$,
\cite{jpa-2003,cjp-2004b}. This means that there is a one-to-one
correspondence between elements of ${\cal H}$ and ${\cal H}_{\rm
phys}$, $\psi\to\psi':=\eta_+^{\frac{1}{2}}\psi$, and the
corresponding physical observables, $A\to
A':=\eta_+^{\frac{1}{2}}A\eta_+^{-\frac{1}{2}}$, such that the
expectation values coincide: $
    \frac{\pbr\psi,A\psi\pkt}{\pbr\psi,\psi\pkt}=
    \frac{\br\psi'|A'\psi'\kt}{\br\psi'|\psi'\kt}$.

We begin our investigation of the state space in pseudo-Hermitian
quantum mechanics by identifying physical states with orthogonal
projection operators onto the corresponding rays. Given a state
vector $\psi\in{\cal H}_{\rm phys}-\{0\}$, the corresponding
orthogonal projection operator $\Lambda_\psi:{\cal H}_{\rm
phys}\to {\cal H}_{\rm phys}$ is a self-adjoint operator
($\Lambda_\psi^\sharp=\Lambda_\psi$) satisfying
$\Lambda_\psi^2=\Lambda_\psi$, $\Lambda_\psi\psi=\psi$, and
$\Lambda_\psi\phi=0$ if $\pbr\psi,\phi\pkt=0$, \footnote{The
self-adjointness and orthogonality of $\Lambda_\psi$ are dictated
by the quantum measurement theory \cite{cjp-2006}.}. These
conditions imply
    \be
    \Lambda_\psi=\frac{|\psi\kt\br\psi|\eta_+}{\br\psi|\eta_+\psi\kt}
    =\frac{|\psi\pkt\pbr\psi|}{\pbr\psi,\psi\pkt},
    \label{lambda}
    \ee
where $|\psi\pkt:=|\psi\kt$ and $\pbr\psi|:=\br\psi|\eta_+$,
\footnote{Another way of deriving (\ref{lambda}) is to use the
Hermitian representation, i.e., demand that
$\Lambda_\psi=\eta_+^{-\frac{1}{2}}\left(\frac{
|\psi'\kt\br\psi'|}{\br\psi'|\psi'\kt}\right)\eta_+^{\frac{1}{2}}$.}.

The inner product $\pbr\cdot,\cdot\pkt$ of ${\cal H}_{\rm phys}$
induces the following inner product on the space of (trace-class)
operators acting in ${\cal H}_{\rm phys}$:
    \be
    (A,B):={\rm tr}(A^\sharp B)={\rm tr}(\eta_+^{-1}A^\dagger
    \eta_+ B),
    \label{trace}
    \ee
where ${\rm tr}(A):=\sum_n \pbr\psi_n,A\psi_n\pkt$ and
$\{\psi_n\}$ is an arbitrary orthonormal basis of ${\cal H}_{\rm
phys}$, \cite{reed-simon}. Using the definition of the inner
product $\pbr\cdot,\cdot\pkt$ and introducing
$\psi_n':=\eta_+^{\frac{1}{2}}\psi_n$ which form an orthonormal
basis in ${\cal H}$, we have ${\rm tr}(A)=\sum_n \br\psi_n|\eta_+
A\psi_n\kt=\sum_n \br\psi'_n|\eta_+^{\frac{1}{2}} A
\eta_+^{-\frac{1}{2}}\psi'_n\kt$. Here the last sum is the usual
trace of $\eta_+^{\frac{1}{2}}A\eta_+^{-\frac{1}{2}}$ which in
view of the cyclic identity for the trace coincides with the trace
of $A$. This shows that ``${\rm tr}$'' is identical with the
conventional trace used in Hermitian quantum mechanics. We also
note that (\ref{trace}) is the unique inner product that for a
given orthonormal basis $\{\psi_n\}$ of ${\cal H}_{\rm phys}$
renders $\{\Lambda_{\psi_n}\}$ orthonormal.

Next, we set $d\Lambda_\psi:=\Lambda_{\psi+d\psi}-\Lambda_\psi$
and define the metric on ${\cal PH}_{\rm phys}$ according to
    \be
    ds^2:={\rm tr}(d\Lambda_\psi^\sharp d\Lambda_\psi).
    \label{defn}
    \ee
After miraculous cancellations of many terms in this lengthy
calculation and using the identity ${\rm
tr}(|\psi\kt\br\phi|)=\br\phi|\psi\kt$, we find
    \bea
    ds^2&=&\frac{2\left[\pbr\psi,\psi\pkt\pbr d\psi,d\psi\pkt-|
    \pbr\psi,d\psi\pkt|^2\right]}{|\pbr\psi,\psi\pkt|^2}\nn\\
    &=&\frac{2\left[\br\psi|\eta_+\psi\kt\br d\psi|\eta_+d\psi\kt-|
    \br\psi|\eta_+d\psi\kt|^2\right]}{|\br\psi|\eta_+\psi\kt|^2}.
    \label{ds2}
    \eea
For the case that ${\cal H}$ is identified as $\C^N$ endowed with
the standard Euclidean inner product, i.e.
$\br\psi|\tilde\psi\kt:=\sum_{n=1}^N z_i^*\tilde z_i$ where
$\psi=(z_1,z_2,\cdots,z_N)$ and $\tilde\psi=(\tilde z_1,\tilde
z_2,\cdots,\tilde z_N)$, Equation~(\ref{ds2}) takes the form
    \be
    ds^2=\sum_{i,j=1}^N g_{ij^*}\,dz_idz^*_j,
    \ee
where
    \be
    g_{ij^*}:=\frac{2\sum_{p,q=1}^N\left[\eta_{pq}\eta_{ji}-
    \eta_{pi}\eta_{jq}\right] z_p^*z_q}{\left(\sum_{m,n=1}^N
    \eta_{mn} z_m^*z_n\right)^2},
    \label{g=}
    \ee
and $(\eta_{ij})$ is the matrix representation of $\eta_+$ in the
standard basis of $\C^N$. If we identify $\eta_+$ with the
identity operator, (\ref{g=}) yields the Fubini-Study metric on
$\C P^{N-1}$, \cite{anandan-aharonov}. To see this in more detail,
consider the case $N=2$. Relabelling the entries of $(\eta_{ij})$
in terms of the real parameters $a,b_1,b_2,c$ as
    \be
    (\eta_{ij})=:\left(\begin{array}{cc} a & b_1+ib_2\\
    b_1-ib_2 & c\end{array}\right),
    \label{eta=}
    \ee
using the homogeneous coordinate $\zeta:=z_2/z_1$ in the patch
where $z_1\neq 0$, i.e., taking $(z_1,z_2)\to(1,\zeta)\to \zeta$
(which is equivalent to setting $z_1=1$ and $z_2=\zeta$), and
letting $\zeta=:x+iy$ with $x,y\in\R$, we obtain
    \be
    ds^2=\frac{2d\,(dx^2+dy^2)}{
    [a+2(b_1x-b_2y)+c(x^2+y^2)]^2},
    \label{fubini-study}
    \ee
where $d:=ac-(b_1^2+b_2^2)={\rm det}(\eta_{ij})$. For
$\eta_{ij}=\delta_{ij}$, i.e., $a=c=1$ and $b_1=b_2=0$,
(\ref{fubini-study}) reduces (up to a factor of 2) to the well-known
formula for the Fubini-Study metric on $\C P^1$ (which is a
two-dimensional sphere of unit diameter)
\cite{eguchi-gilkey-hanson}.

Equations~(\ref{lambda}) and (\ref{ds2}) show that because the
inner product of the physical Hilbert space depends on the
Hamiltonian, so do the orthogonal projection operators
representing the states and the metric on the state space. This is
the root of the subtleties of the Brachistochrone problem in
pseudo-Hermitian quantum mechanics. There is a fundamental
difference between this problem and its Hermitian analog. Its
proper formulation as a standard variational problem must include
fixed (Hamiltonian-independent) choices for the initial and final
states (as opposed to unobservable state vectors) as well as for
the geometry of the space in which one minimizes the travel time.
One can apply the argument of \cite{brody-hook} and use the
results of \cite{anandan-aharonov} to identify the travel time
with a multiple of the distance travelled by the evolving state in
the state space ${\cal PH}_{\rm phys}$. This would make the lower
bound on the travel time proportional to the geodesic distance
between the initial and final states.

In the treatment of the problem offered in \cite{bender-prl-2007},
both the boundary conditions and the very notion of distance
depend on the choice of the Hamiltonian. Therefore, it is not
clear whether the result corresponds to arbitrarily close initial
and final states or arbitrarily short travel times for distant
states. To conclude that one may achieve ``faster than Hermitian
quantum mechanics'' evolutions, one must consider initial and
final states with a fixed distance and consider whether one can
find evolutions that take less time than the lower bound set by
Hermitian quantum mechanics. In the remainder of this paper we
prove that this is indeed impossible \footnote{This is a stronger
result than showing the impossibility of arbitrarily short travel
times for distant initial and final states. The latter may be
simply established using the fact that the travel time is
proportional to the geodesic distance defined by the metric
(\ref{g=}).};

\noindent {\bf Theorem:} \emph{The lower bound on the travel time
(upper bound on the speed) of unitary evolutions is a universal
quantity independent of whether the evolution is generated by a
Hermitian or non-Hermitian Hamiltonian.}

To prove this theorem consider an arbitrary non-Hermitian
Hamiltonian operator $H$ that generates a unitary time-evolution
in the Hilbert space ${\cal H}_{\rm phys}$ defined by a metric
operator $\eta_+$. Let $\psi_I$ and $\psi_F=U(\tau)\psi_I$ be the
initial and final state vectors, where $U(t):=e^{-\frac{it
H}{\hbar}}$ is the evolution operator for $H$ and $\tau\in\R^+$ is
the travel time. Then the evolution operator $u(t):=e^{-\frac{it
h}{\hbar}}$ for the Hermitian Hamiltonian
$h:=\eta^{\frac{1}{2}}H\eta^{-\frac{1}{2}}$, which defines a
dynamics in the Hilbert space ${\cal H}$, evolves
$\psi_I':=\eta^{\frac{1}{2}}\psi_I$ into
$\psi_F':=\eta^{\frac{1}{2}}\psi_F=u(\tau)\psi'_I$ in time $\tau$.
This follows from the fact that $U(t)$ is $\eta_+$-pseudo-unitary,
i.e., $U(t)^{-1}=U(t)^\sharp=\eta_+^{-1}U(t)^\dagger\eta_+$,
\cite{jmp-2004}. Next, recall that
    \be
    \eta_+^{\frac{1}{2}}:{\cal H}_{\rm phys}\to{\cal H}
    \label{unitary-x}
    \ee
is a unitary operator \cite{jpa-2003,cjp-2004b}, i.e., for all
$\psi,\phi\in{\cal H}_{\rm phys}$, $\pbr\psi,\phi\pkt
=\br\eta_+^{\frac{1}{2}}\psi| \eta_+^{\frac{1}{2}}\phi\kt$. In
view of this relation and (\ref{lambda}), it is not difficult to
show that while the state $\Lambda_I$ evolves into $\Lambda_F$ in
${\cal PH}_{\rm phys}$, the state
$\Lambda'_I:=\eta^{\frac{1}{2}}\Lambda_I\eta^{-\frac{1}{2}}$
evolves into
$\Lambda'_F:=\eta^{\frac{1}{2}}\Lambda_F\eta^{-\frac{1}{2}}$ in
${\cal PH}$. Because the travel times are identical ($=\tau$), the
speed of the evolution generated by $H$ will be different from
that generated by $h$ if and only if the length of the curve
$\Lambda(t)$ joining $\Lambda_I$ to $\Lambda_F$ in ${\cal PH}_{\rm
phys}$ is different from that of the curve $\Lambda'(t)$ joining
$\Lambda'_I$ to $\Lambda'_F$ in ${\cal PH}$. The optimal-time
evolution for $H$ corresponds to the case that $\Lambda'(t)$ is a
geodesic in ${\cal PH}_{\rm phys}$. To prove the above theorem,
therefore, it is sufficient to show that the geodesic distance
between $\Lambda_I$ and $\Lambda_F$ in ${\cal PH}_{\rm phys}$ is
identical with the geodesic distance between $\Lambda'_I$ and
$\Lambda'_F$ in ${\cal PH}$. This follows from the fact that the
diffeomorphism $f:{\cal PH}_{\rm phys}\to{\cal PH}$ that is
induced by the unitary transformation (\ref{unitary-x}), namely
    \be
    f(\Lambda):=\eta^{\frac{1}{2}}\Lambda\eta^{-\frac{1}{2}},
    \label{f=}
    \ee
pulls back the Fubini-Study metric on ${\cal PH}$ to the metric
(\ref{g=}) on ${\cal PH}_{\rm phys}$. In other words \emph{$f$ is
an isometry}. Probably the simplest way of showing this is to use
(\ref{defn}), (\ref{f=}), and the cyclic identity for the trace to
establish
    \be
    ds^2={\rm tr}(f(d\Lambda_\psi)^\dagger f(d\Lambda_\psi))
    \ee
and note that the Fubini-Study metric corresponds to $ds^2={\rm
tr}(d{\Lambda'_{\psi}}^{\!\dagger} d\Lambda'_{\psi})$ where
$\Lambda'_{\psi}:=\frac{|\psi\kt\br\psi|}{\br\psi|\psi\kt}$.

Because the distance between $\Lambda_I$ and $\Lambda_F$ in ${\cal
P H}_{\rm phys}$ is the same as the distance between $\Lambda'_I$
and $\Lambda'_F$ in ${\cal P H}$, and because given a
non-Hermitian Hamiltonian $H$ that evolves $\Lambda_I$ into
$\Lambda_F$ in time $\tau$ the Hermitian Hamiltonian $h$ evolves
$\Lambda'_I$ into $\Lambda'_F$ in the same time $\tau$, the
evolution speed for $H$ is identical to that of $h$. In
particular, there is no advantage of using a non-Hermitian
Hamiltonian $H$ as far as the lower bound on $\tau$ is concerned.
This argument shows that a vanishing lower bound corresponds to
arbitrarily close initial and final states; it can never be
achieved for distant initial and final states. For the case of
antipodal initial and final states \cite{Vaidman}, which is
directly relevant to the problem of constructing unitary NOT
operations in quantum computation \cite{Margolus}, we can verify
this statement directly.

First we note that without loss of generality we can confine our
attention to two-level systems. If we represent the initial state
$\Lambda_I$
by the state vector $\psi_I=\left(\begin{array}{c}1\\
0\end{array}\right)$, the antipodal (final) state $\Lambda_F$,
that satisfies $\Lambda_F\Lambda_I=\Lambda_I\Lambda_F=0$, will be
represented by a state vector of the form $\psi_F=\nu
\left(\begin{array}{c}\beta\\-a\end{array}\right)$, where $\nu$ is
an arbitrary nonzero normalization constant,
$\beta:=b_1+ib_2=\eta_{12}$, and we have enforced the condition
$\br\psi_F|\eta_+\psi_I\kt=\pbr\psi_F,\psi_I\pkt=0$ and used
(\ref{eta=}). The initial and final states have the form $
\Lambda_I=\left(\begin{array}{cc}1&\frac{\beta}{a}\\
0 & 0\end{array}\right)$,
$\Lambda_F=\left(\begin{array}{cc}0&-\frac{\beta}{a}\\
    0 & 1\end{array}\right)$,
respectively. This calculation shows that fixing the initial and
final states puts a restriction on the choice of the metric
operator $\eta_+$ and consequently the allowed Hamiltonian
operator $H$. One can show that for $\psi_F=\left(\begin{array}{c}0\\
1\end{array}\right)$, the metric operator $\eta_+$ must be
diagonal and $S_z=\frac{\hbar}{2}\left(
\begin{array}{cc}1&0\\
0 & -1\end{array}\right)$ satisfies $S_z^\sharp=S_z$, i.e., it is
an observable. For general non-diagonal $\eta_+$, $S_z$ fails to
be an observable; it does not describe the spin along the $z$-axis
and its eigenstates are not the states of definite spin along the
$z$-axis.

The general form of an $\eta_+$-pseudo-Hermitian two-level
Hamiltonian for the most general $\eta_+$ is given in
\cite{tjp-2006}. To find the optimal-time evolution for the above
boundary conditions, we can pursue three different approaches: (i)
We can use the results of \cite{tjp-2006} to determine the general
form of $H$ for a given value of $\beta/a$ and fix the remaining
degrees of freedom in $H$ by minimizing the travel time; (ii) We
can follow the approach of \cite{brody-hook} and compute the
minimum travel time by evaluating the geodesic distance between
$\Lambda_I$ and $\Lambda_F$ using the metric (\ref{fubini-study});
(iii) We can map the problem to the one in the standard Hilbert
space ${\cal H}$ and use the known results for Hermitian
Hamiltonians. Here we employ (iii), because it is more
straightforward.

Let $\Lambda'_I$ and $\Lambda'_F$ be the corresponding states in
${\cal PH}$. Then a quick calculation shows that $\Lambda'_I$ and
$\Lambda'_F$ are also antipodal;
$\Lambda'_I\Lambda'_F=\eta^{\frac{1}{2}}\Lambda_I\Lambda_F
\eta^{-\frac{1}{2}}=0$. Therefore, as implied by the above
theorem, the lower bound on the travel time $\tau$ is identical
with $\hbar\pi/|E_1-E_2|$ where $E_1$ and $E_2$ are the
eigenvalues of $h$, \cite{brody-hook}. Note that because $h$ and
$H$ are isospectral, $E_1$ and $E_2$ are also eigenvalues of $H$.

This calculation confirms the statement of the above theorem for
the case that the initial and final states are antipodal states.
It shows that for these boundary conditions the bound obtained for
Hermitian Hamiltonians also applies for admissible non-Hermitian
Hamiltonians. Therefore, non-Hermitian Hamiltonians (that are
capable of generating unitary time-evolutions) do not offer any
advantage in performing a faster unitary NOT-operation. This
conclusion cannot be avoided unless one sacrifices unitarity. Note
that the hypothetical setups that involve switching between
Hermitian and non-Hermitian Hamiltonians at different times
\cite{bender-prl-2007} would require a time-dependent metric
operator which in turn violates unitarity \cite{plb-2007}.
Therefore such scenarios cannot be used to undermine the general
applicability of the above theorem.

In this article we examined the structure of the projective space
${\cal PH}_{\rm phys}$ of physical states in pseudo-Hermitian
quantum mechanics. We derived the form of the natural metric
tensor on ${\cal PH}_{\rm phys}$ and showed that as a Riemannian
manifold it is isometric to the projective Hilbert space ${\cal
PH}$ of Hermitian quantum mechanics. Furthermore, we demonstrated
that the time-evolution in ${\cal PH}_{\rm phys}$ that is
determined by a diagonalizable non-Hermitian Hamiltonian with a
real spectrum has a mirror image in the usual state space ${\cal
PH}$ of Hermitian quantum mechanics. This is indeed a
manifestation of the equivalence of pseudo-Hermitian (and in
particular ${\cal PT}$-symmetric) quantum mechanics with Hermitian
quantum mechanics. A direct consequence of this equivalence is
that physical quantities cannot differentiate between
pseudo-Hermitian and Hermitian quantum mechanics. Pseudo-Hermitian
quantum mechanics can only be useful as a technical tool
particularly for dealing with systems with an infinite-dimensional
Hilbert space where unlike $H$ the equivalent Hermitian
Hamiltonian $h$ is generically a complicated nonlocal operator.
Typical examples are the imaginary cubic potential, $H=p^2+ix^3$,
and the delta function potential with a complex coupling,
$H=p^2+z\delta(x)$, where $z\in\C$, \cite{jpa-2006ab}.

\noindent {\bf Acknowledgment:} I wish to thank Zafer Gedik for
bringing reference \cite{Margolus} to my attention.

\ed